%% file: bilayers_main.tex
\documentclass{article}

\usepackage[right= 0.75in, left= 0.75in, top=1.05in, bottom=1.15in]{geometry}

\usepackage{lineno,hyperref,natbib}
\modulolinenumbers[5]

\usepackage{color}

\usepackage{amsmath}
\usepackage{amsfonts}
\usepackage{bm} %boldmath
\usepackage{booktabs} % for much better looking tables
\usepackage{array} % for better arrays (eg matrices) in maths
\usepackage{paralist} % very flexible & customisable lists (eg. enumerate/itemize, etc.)
\usepackage{verbatim} % adds environment for commenting out blocks of text & for better verbatim
\usepackage{subfig} % make it possible to include more than one captioned figure/table in a single float
\usepackage[utf8]{inputenc}
\usepackage[affil-it,auth-sc]{authblk}
\usepackage{graphicx}
\usepackage{amsmath}
\newcommand{\mb}[1]{\mathbf{#1}} % bold math

\newcommand{\mc}[1]{\mathcal{#1}} % mathcal
\author[1]{No\`{e} Caruso\thanks{Authors contributed equally.}}
\author[2]{Aleksandar Cvetkovi\'{c}$^*$}
\author[1]{Alessandro Lucantonio}
\author[1]{Giovanni Noselli}
\author[1,2]{Antonio DeSimone\thanks{Corresponding author -- e-mail address:\,\texttt{desimone@sissa.it}.}}
\affil[1]{\small{SISSA--International School for Advanced Studies, via Bonomea 265, 34136 Trieste, Italy.}}
\affil[2]{GSSI--Gran Sasso Science Institute, viale Francesco Crispi 7, 67100 L'Aquila, Italy}

\title{Spontaneous morphing of equibiaxially pre-stretched elastic bilayers: \\ the role of sample geometry}

\date{}

\begin{document}
	
\maketitle
	
\begin{abstract}
An elastic bilayer, consisting of an equibiaxially pre-stretched sheet bonded to a stress-free one, spontaneously morphs into curved shapes in the absence of external loads or constraints. Using experiments and numerical simulations, we explore the role of geometry for square and rectangular samples in determining the equilibrium shape of the system, for a fixed pre-stretch. We classify the observed shapes over a wide range of aspect ratios according to their curvatures and compare measured and computed values, which show good agreement.  In particular, as the bilayer becomes thinner, a bifurcation of the principal curvatures occurs, which separates two scaling regimes for the energy of the system. We characterize the transition between these two regimes and show the peculiar features that distinguish square from rectangular samples. The results for our model bilayer system may help explaining morphing in more complex systems made of active materials. 
\end{abstract}

\input{bilayers_intro.tex}

\input{bilayers_theory.tex}

\input{bilayers_experiment.tex}

\input{bilayers_discussion.tex}

\input{bilayers_conclusions.tex}

\section*{Acknowledgements}
The authors gratefully acknowledge financial support from the European Research Council through AdG-340685 ``MicroMotility''. We thank prof.~E.~Sharon for valuable discussions.

\end{document}

%% file: bilayers_intro.tex
\section{Introduction}

This paper is part of a broader programme to try and acquire a thorough understanding of morphing in elastic structures made of active materials through the use of model systems. In particular, we are interested in model systems  that are  easily manufactured in a laboratory and easy to  reproduce in numerical simulations based on the finite element method. Our plan is to use the results obtained with the study of the model system to address similar questions of shape programming in more complicated physical systems. These are systems that undergo heterogenous spontaneous deformations as a consequence of, say,  phase transformations, swelling due to the absorption of a liquid, or in response to either thermal or electromagnetic stimuli.

The study of shape changes induced by a heterogeneous response to external stimuli in thin structures has a long history. Among the classical contributions regarding the thermal buckling of a bimetallic strip or disc, we refer to \citep{timoshenko25} and \citep{freund2000}.
{\color{black}Variants of this problem are the study of thermal-induced delamination and blistering of solid thin film coatings deposited over solid substrates (see, e.g. \citep{gioia2002,gioia97}), or morphing through the use of martensitic thin films (e.g., made of shape-memory-alloys \citep{Bhat_1999}) or of semiconductor bilayers deposited using molecular beam epitaxy \citep{Danescu_2013}.}
While predicting and controlling a shape change of a given system is interesting in itself, these shape changes can be exploited to perform useful functions like, for example, the use of bimetallic strips as thermostats \citep{timoshenko25}, pumping fluid in a micro-fluidic circuit \citep{Bhat_1999}, or for the locomotion of soft, bioinspired micro-robots \cite{gidoni15,tatone13,noselli14}.

The problem of exploiting material heterogeneities to induce controlled shape changes as a response to an external, spatially homogeneous stimulus, has received considerable attention in recent years.   
Often, shape is controlled by generating curvature  in thin structures. In the classical application of bimetallic strips, the difference between thermal expansion coefficients of the two metals generates differential strains along the thickness direction, and hence elastic curvature. 
A different principle exploits  spontaneous curvature arising from strain variations in the mid-plane of a plate. This mechanism is based on the Gauss' Theorema Egregium \citep{arroyo2014,klein2007}. Both curvature-generating principles can be implemented in polymer gels \cite{kim2012,lucantonio2014,lucantonio2016,na2015,pezzulla2015,pezzulla2016} and in Liquid Crystal Elastomers, an interesting polymeric (hence soft) active material which is receiving increasing attention \citep{agostiniani,Ago_DeS_Kou,Aha_Sha_Kup,Bhat_Mod_War,Mostajeran}. 
Some of the interesting spontaneous morphing behaviour of Liquid Crystal Elastomers can be reproduced in pre-stretched bilayers, made of simple (i.e., not-active) elastic materials. It is shown in \citep{agostiniani2017,Bartels,desimone2017} that, by ``engineering'' the prestretch in each of the two layers of a bilayer one can reproduce different structural models, such as bistable shells and shells with very low stiffness with the respect to twisting, as discussed in \citep{guest2006} and \citep{guest2011}, respectively. Moreover, by shaping the mid-plane of the bilayer in the form of a long, narrow rectangle, and varying the angle between the long axis of the rectangle and the uniaxial prestretch direction in one of the layers, one can produce tunable helical ribbons as in \citep{chen2011}.

Inspired by \citep{chen2011}, we study an elastic bilayer where an equibiaxially pre-stretched sheet is bonded to a stress-free one. Free from constraints or applied loads, such a system spontaneously relaxes to a curved equilibrium by developing differential strains along the thickness. Previous studies have focused on the role of strain mismatch in determining the equilibrium shape, which was induced by different thermal expansion coefficients \citep{mansfield1962,timoshenko25}, swelling factors \citep{pezzulla2016} or pre-stretches \citep{freund2000}, or by combining an electrically-active layer with a passive one \citep{alben2011}. 
Here, by combining experiments on latex samples and numerical simulations, we explore instead the role of sample geometry for a fixed pre-stretch.  In particular, as the width-to-thickness ratio varies, the system shows a bifurcation of the equilibrium path, which we track by measuring the principal curvatures. Interestingly, we find that the transition is continuous for squares and discontinuous for rectangles, in terms of the difference between the principal curvature. Moreover, we report a shift in the onset of the transition towards higher values of the width-to-thickness ratio for squares as compared to rectangular samples. More generally, we provide a quantitative comparison between theory and experiments over a wide range of aspect ratios and a classification of the shapes, which have not been reported previously.

The paper is organized as follows. In Section~\ref{sec:methods} we introduce the theoretical model describing the elasticity of the bilayer, the numerical procedures for the computation of the equilibrium paths, and the experimental protocol for the preparation of latex bilayers. In Section~\ref{sec:results} we present and compare experimental and numerical results concerning the curvatures of the observed shapes for various aspect ratios. Finally, in Section~\ref{sec:conclusions} we draw some conclusions and indicate possible future directions.

%% file: bilayers_theory.tex
\section{Materials and Methods}
\label{sec:methods}

We consider a bilayer of length $L$, width $W$ and total thickness $H'$ consisting of two thin, elastic sheets of the same material and bonded together (Figure~\ref{fig:1}a). The bottom sheet has been pre-stretched before bonding, such that, in the absence of applied tractions or constraints, the flat configuration is not an equilibrium. Thus, the system spontaneously deforms into a three-dimensional configuration by partially releasing the elastic energy stored in the pre-stretched layer. To characterize the emergent equilibrium shapes, we establish in the following a theoretical model accounting for the elasticity of the system and describe the computational strategy for its solution.

\subsection{Theoretical modeling and numerical procedures}

\begin{figure}
    \centering
    \includegraphics[scale=0.22]{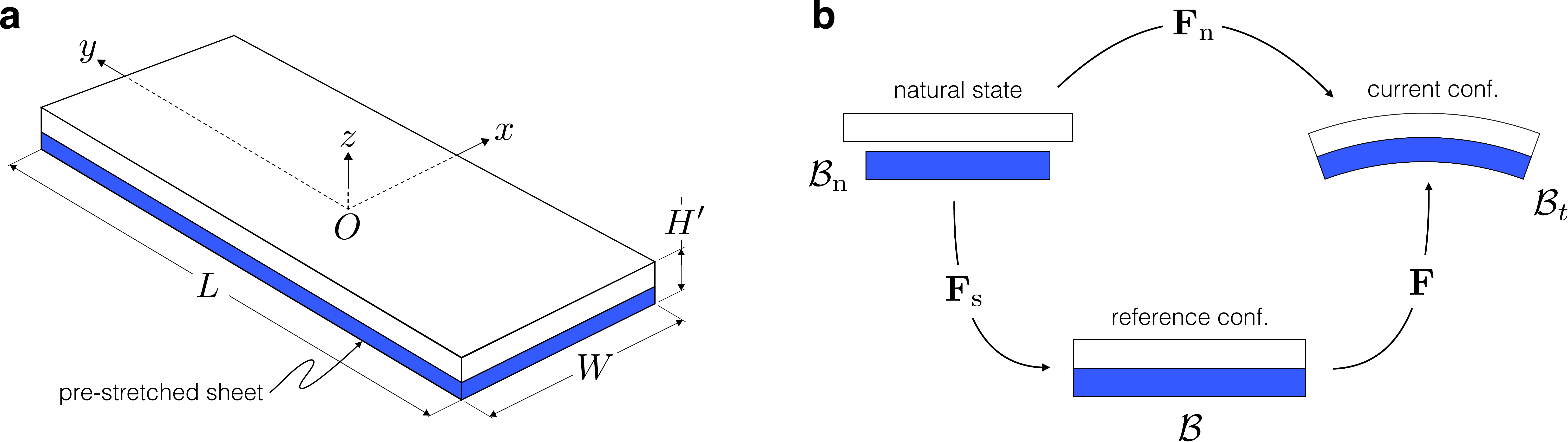}
    \caption{(a) Sketch of the reference configuration of the system and the global reference frame. The colored sheet is pre-stretched equibiaxially in the plane $xOy$. (b) Decomposition of the deformation process from the natural state to the current configuration.}
    \label{fig:1}
  \end{figure}

Following a modeling approach close to that reported in \cite{huang2012}, we  account for the pre-stretch of the bottom layer by means of the multiplicative decomposition of the deformation gradient $\mb{F}_{\rm n}$ from the natural state $\mc{B}_{\rm n}$ of the system into its current equilibrium configuration $\mc{B}_t$ as $\mb{F}_{\rm n} = \mb{F}\mb{F}_{\rm s}$, where
  \begin{itemize}
    \item $\mb{F}_{\rm s}$ is the deformation gradient describing the application of the pre-stretch and the bonding of the sheets starting from the natural state, where the sheets are separate and undeformed;
    \item $\mb{F}=\mb{I}+\nabla\mb{u}$ is the deformation gradient from the reference configuration to the current configuration, with $\mb{u}$ the corresponding displacement field.
  \end{itemize}
This decomposition and the relations among the configurations are schematically depicted in Figure~\ref{fig:1}b. Consistent with the experiments, we assume that the bottom sheet is pre-stretched equibiaxially in the $xOy$ plane while allowing for thickness contraction, so that the deformation gradient $\mb{F}_{\rm s}$ is represented as
\begin{equation}
	\mb{F}_{\rm s} = \lambda (\mb{e}_1\otimes\mb{e}_1 + \mb{e}_2\otimes\mb{e}_2) + \lambda_3 \mb{e}_3 \otimes \mb{e}_3\,, \label{eq:1}
\end{equation}
where $\lambda$ and $\lambda_3$ the in-plane and out-of-plane stretches, respectively. 
The current equilibrium configuration is determined by the balance of forces that, in the absence of inertia and external loads, reads
  \begin{equation}
    \int_{\mathcal{B}}\mathbf{S}\cdot\nabla\mathbf{\tilde{u}} \,\mathrm{dV} = 0 \,, \label{eq:10}
  \end{equation}
where $\tilde{\mb{u}}$ is a kinematically admissible test displacement field and $\mb{S}$ is the Piola-Kirchhoff stress tensor in the reference configuration.
As for the constitutive equations, we model the sheets as compressible neo-Hookean materials, characterized by the following representation of the free energy density $\psi_{\rm n}$ in the natural state \citep{bonetwood}
  \begin{equation}
    \psi_{\rm n}\left(\mb{F}_{\rm n}\right) = \frac{G}{2}\left(\mb{F}_{\rm n} \cdot \mb{F}_{\rm n} - 3 - 2\,\log J_{\rm n} \right) + \frac{\Lambda}{2}(\log J_{\rm n})^{2} \,, \label{eq:11}
  \end{equation}
where $J_{\rm n}  = \det \mb{F}_{\rm n}$, whereas $G$ and $\Lambda$ are the Lamé moduli of the material.  From this expression the energy density $\psi$ per unit reference volume may be readily computed as $\psi(\mb{F}) = \psi_{\rm n}(\mb{F}\mb{F}_{\rm s})/J_{\rm s}$, with $J_{\rm s} = \det \mb{F}_{\rm s}$, and the corresponding referential stress is given by
\begin{align}
{\color{black}
\mb{S} = \frac{\partial \psi}{\partial \mb{F}} = \frac{G}{J_{\rm s}}\left(\mb{F}\mb{F}_{\rm s}\mb{F}_{\rm s}^{\rm T}-\mb{F}^{-\rm T}\right)+\frac{\Lambda}{J_{\rm s}} \log\,(J J_{\rm s})\mb{F}^{-\rm T} , 
}
\label{eq:consteq}
\end{align}
where $J = \det \mb{F}$. Upon noting that the surfaces of the bottom layer with normal $\pm\,\mb{e}_3$ are traction-free as the pre-stretch is introduced, the principal stretch $\lambda_3$ of eq.~\eqref{eq:1} can be computed by exploiting the boundary condition $\mb{Se}_3 \cdot \mb{e}_3 = 0$, leading to
\begin{align}
G(\lambda_3^2-1) + \Lambda \log\,(\lambda^2 \lambda_3) = 0\,. \label{eq:prestretch}
\end{align}

We solved eq.~\eqref{eq:10} together with the constitutive equations~\eqref{eq:11}-\eqref{eq:consteq} numerically by the finite element method, the displacement field $\mb{u}$ being the only unknown of the problem, which we discretized using quadratic shape functions. Specifically, the commercial software COMSOL Multiphysics (version 5.2a) was employed for the implementation, and a structured mesh of hexahedral elements was used, locally refined at the boundaries to capture boundary layers. To reduce computational costs, symmetry conditions were enforced on the two cross-sections of the body in the $xOz$ and $yOz$ planes and zero vertical displacement was imposed at the origin $O$ to remove rigid body motions. The simulations were set using physical parameters of $E = 1.4\, \rm{MPa}$ and $\nu = 0.49$ for the Young's modulus and Poisson's ratio, respectively, which correspond to the material parameters of the latex sheets used in the experiments. Standard relations from elasticity theory allow then to compute the Lamé moduli as: $G = E/2(1+\nu), \Lambda = E\nu/(1+\nu)(1-2\nu)$. The total thickness of the unstretched sheets in the natural state was set to $H=0.7\ \rm{mm}$ and the in-plane stretch to $\lambda = 1.075$. Notice that the total thickness of the bilayer after application of the pre-stretch is $H' = (1+\lambda_3)H/2$.  

As usual in non-linear problems, solving the weak formulation of eq.~\eqref{eq:10} required the use of a continuation strategy to attain the desired pre-stretch. Specifically, the amount of pre-stretch was ramped linearly in time, which here plays the role of a load parameter, so that at $t=0$ the reference configuration coincides with the unstretched equilibrium.  To avoid re-meshing  of the reference configuration due to the change of thickness with pre-stretch as dictated by eq.~\eqref{eq:prestretch}, the \(z\)--coordinate in the pre-stretched layer was non-dimensionalized by a factor of $\lambda_{3}$ and appropriate scaling of the governing equations was performed, as well, so that the problem could be solved on a computational domain of fixed thickness $H$.  At every time step, the non-linear algebraic problem stemming from the discretization of eq.~\eqref{eq:10} was solved using Newton's method.  

To explore the landscape of equilibrium shapes obtainable by varying the aspect ratios of the bilayer for a fixed pre-stretch, we performed a parametric sweep in $L/W$ and $W/H$. Above a certain threshold value of $W/H$, a bifurcation in the principal curvatures of the equilibrium configuration occurs. 
By this we mean that, for small values of $W/H$, equilibrium shapes are characterized by principal curvatures differing only slightly (in fact, the two principal curvatures are exactly equal for square samples, $L/W=1$). Above a threshold value of $W/H$, equilibrium shapes show two easily distinguishable, different principal curvatures.

Near the bifurcation the system is very sensitive to variations in pre-stretch
and exhibits abrupt curvature changes. We overcame the convergence problems of the numerical procedure related to these sharp changes by transforming the quasi-static continuation problem into a pseudo-dynamic problem with the addition of a viscous damping term into the weak form \eqref{eq:10}. The variable-order (1-5) backward differentiation formula (BDF) was chosen as a time-stepping algorithm. The damping coefficient was tuned in such a way that  the kinetic energy was negligible with respect to the elastic energy in the current configuration. Finally, to select a post-bifurcation equilibrium path, we employed two pre-stretch continuation paths that consisted, respectively, in increasing first the pre-stretch in the \(x\)--direction and then in the \(y\)--direction, and vice-versa.

%% file: bilayers_experiment.tex
\subsection{Experimental setting}
\label{sec:exp}

\begin{figure}[h!]
\centering
	\includegraphics[scale=0.36]{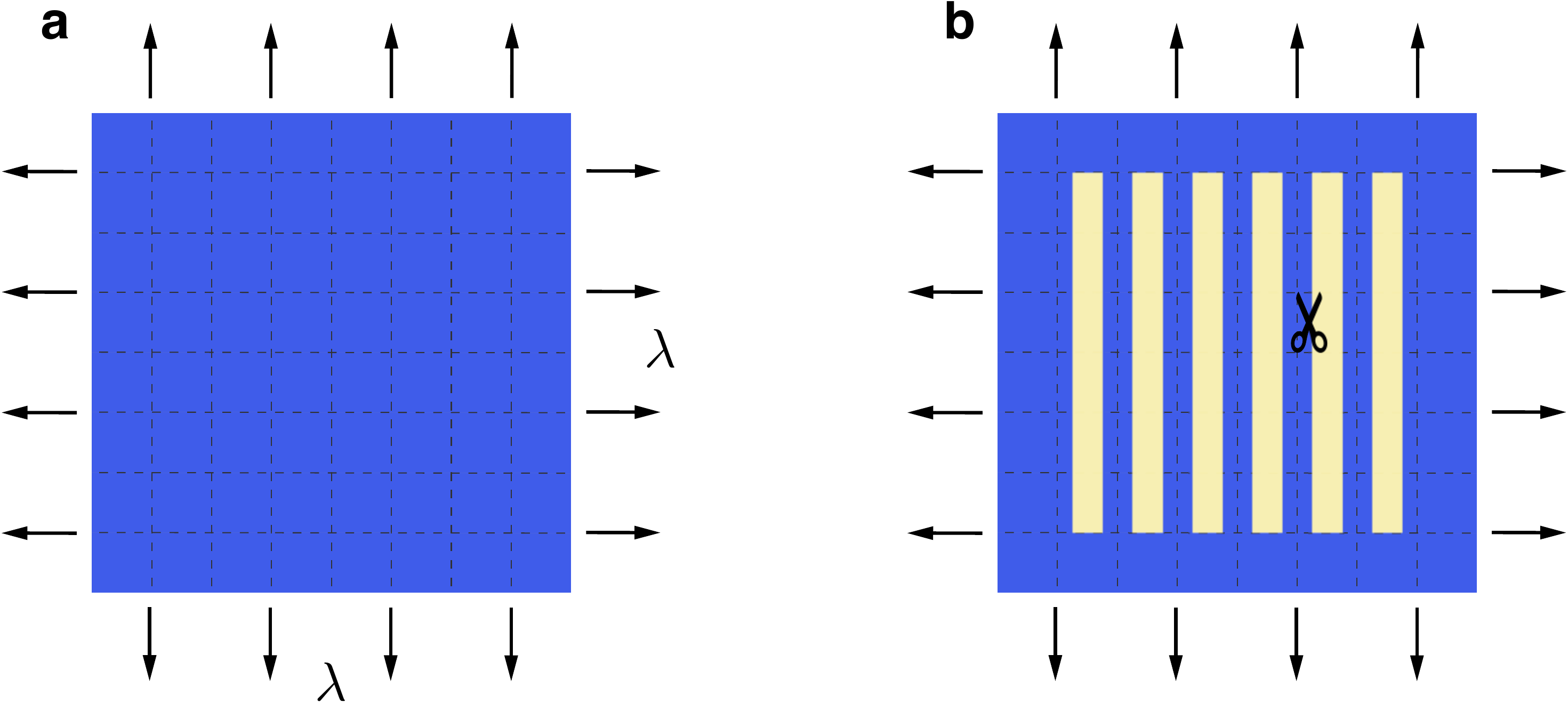}
	\caption{Sketch of the procedure employed to realize pre-stretched bilayer samples. (a) Equibiaxial stretching of a blue latex sheet by a controlled amount $\lambda = 1.075$. (b) Gluing of white unstretched strips of latex and cutting out of bilayer strips.}
	\label{fig:exp}
\end{figure}

We employed a fairly simple experimental setting to obtain lab samples. Latex sheets of thickness $0.35\,\text{mm}$ (from Modulor GmbH, models 4500328-002:018) and liquid, spray adhesive (from Extrema International Srl, model ext-pf-075) were all the essential ingredients we needed.
A custom-made apparatus was designed and constructed for the stretching of the latex sheets. This consisted in four, independent aluminium bars of \lq C-shaped' cross section and included 3D-printed, custom-designed grips that were used to pull on a sheet of latex. A colour coding was employed to indicate whether a sheet was pre-stretched or not. Specifically, blue was chosen to distinguish the pre-stretched bottom layer from the white latex strips that remained unstretched, see the sketch of Figure~\ref{fig:exp}b.

Using the same setting as for the experiments, we first measured the Poisson's ratio of latex. We drew a square grid on a latex sample and then stretched it in one direction, while allowing for its lateral contraction. Thereby, the squares of the grid elongated along the stretching direction and shrunk in the orthogonal one, thus deforming into rectangles.
{\color{black}By measuring the ratio of the lateral to longitudinal strain from the squares sides, we estimated a Poisson's ratio of $\nu = 0.49$. This is consistent with values available from the literature~\citep{chen2011}, which also reports a Young's modulus for latex of $1.4\,\text{MPa}$.}
As for the preparation of bilayer samples, we employed the following procedure. First, we drew a square grid on a sheet of blue latex. By setting the 3D-printed grips at the distance of $50\,\text{mm}$ from one another and by using the custom-made apparatus, we stretched the sheet equibiaxially by a prescribed amount. In particular, a stretch of 1.075 was chosen for the present study. During this step, the grid was exploited to control the magnitude of the pre-stretch and to make sure that the latex sheet was in a state of homogeneous deformation (Figure~\ref{fig:exp}a). Next, we glued 50\,mm wide strips of white latex on top of the pre-stretched layer, and let the spray adhesive dry for three hours to achieve optimal bond strength. Finally, we cut out bilayer strips of white and blue latex and from these obtained a range of lab samples of various dimensions (Figure~\ref{fig:exp}b). As regards the total thickness of the samples, this was dictated by the procedure just described. From samples of unstretched glued bilayer, we measured an average thickness of $0.7\,\text{mm}$, being negligible the contribution from the adhesive film. Depending on their length to width ratio, $L/W$, and width to thickness ratio, $W/H$, the samples deformed into different shapes, as described in the following section.

%% file: bilayers_discussion.tex
\section{Results and discussion}
\label{sec:results}

\begin{figure}[!b]
\centering
\includegraphics[scale=0.3]{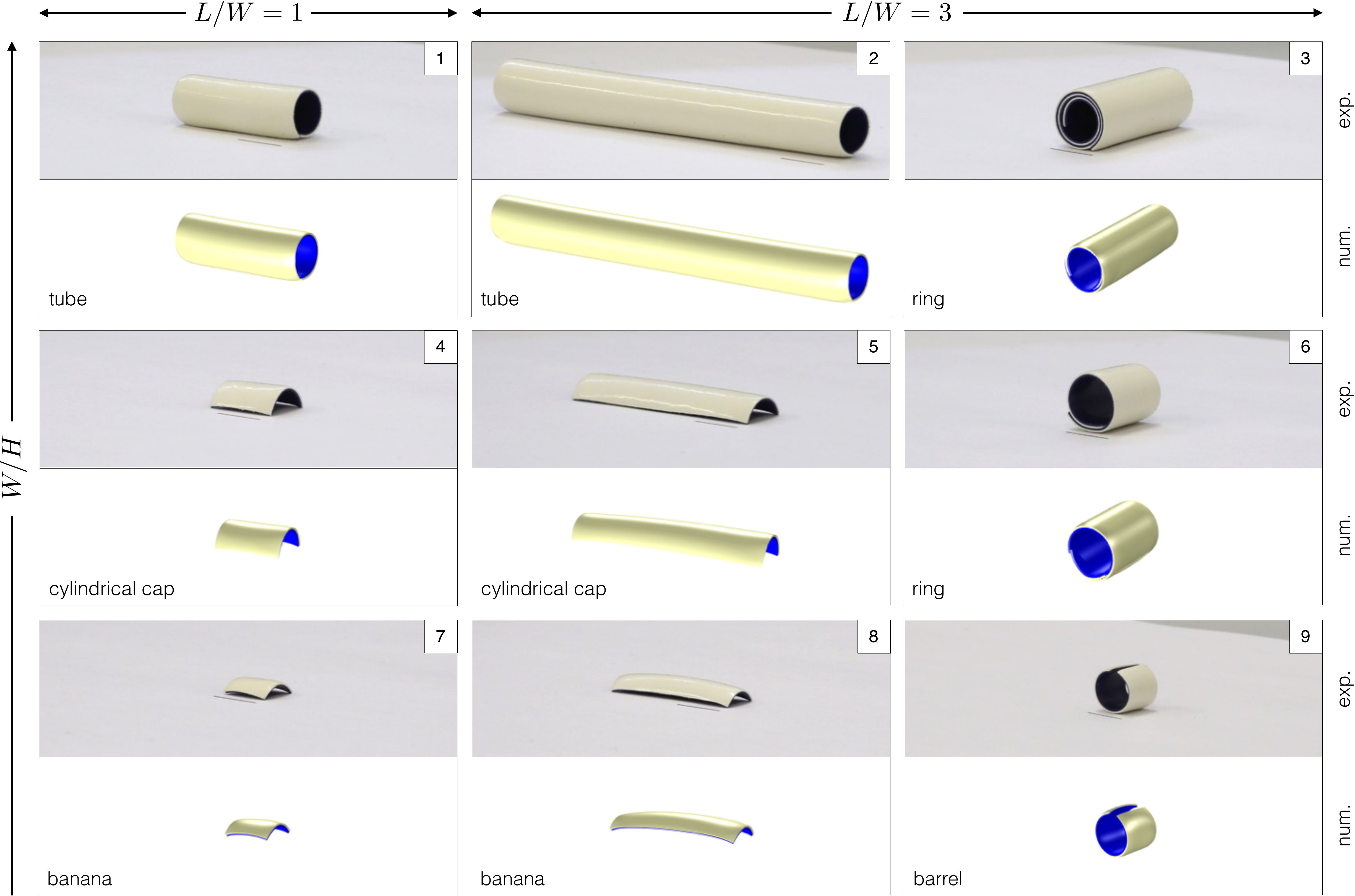}
\caption{Shapes observed in experiments and computed from simulations for $14\leq W/H \leq 43$ and for $L/W=1$ (first column), $L/W=3$ (second and third columns). Numbering of the samples conforms to the measurements of the curvatures reported in Table~\ref{tab:curvatures}. The scale bar corresponds to 10\,mm.}
\label{fig:shapes1}
\end{figure}

\begin{figure}[!t]
\centering
\includegraphics[scale=0.32]{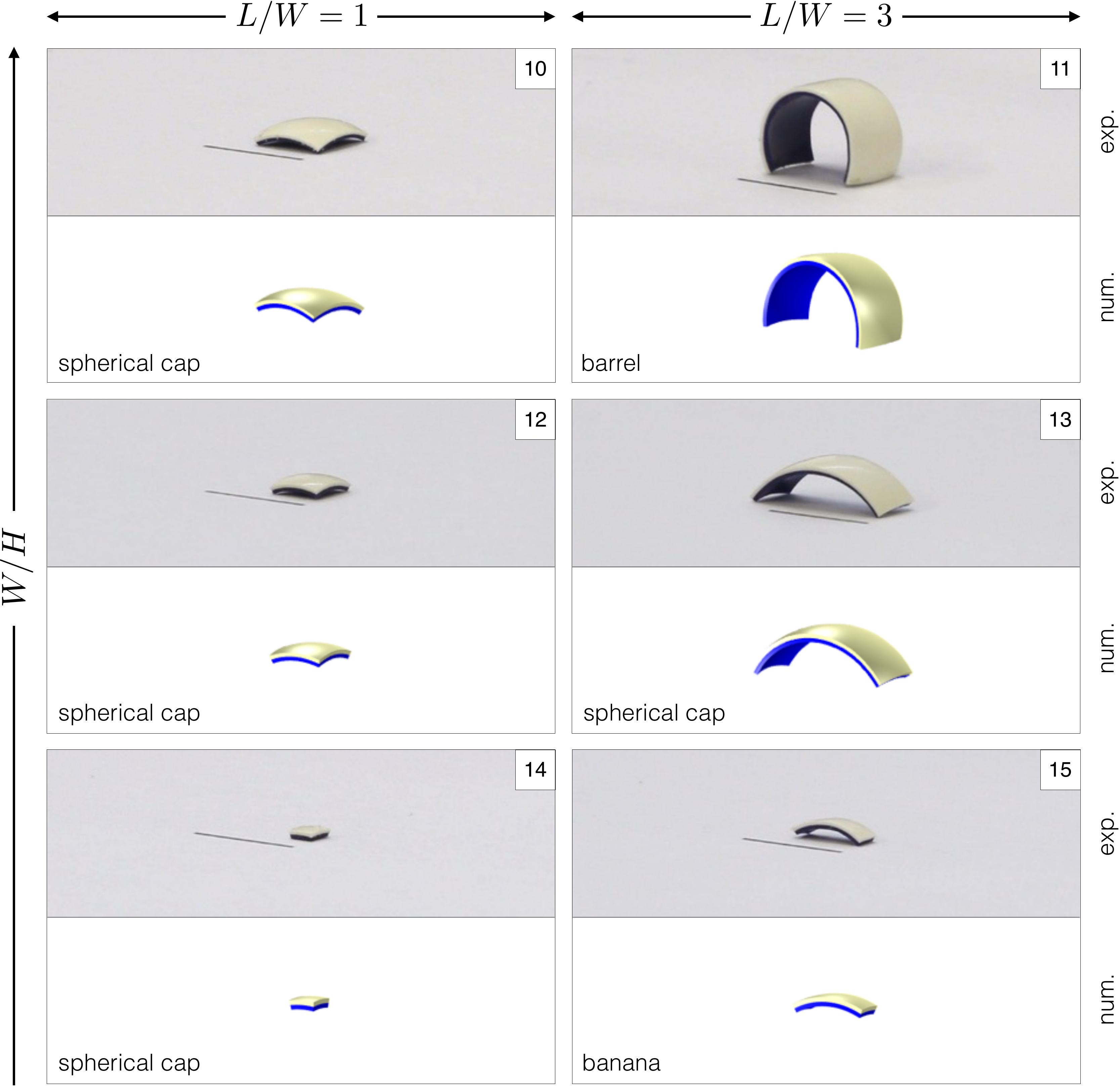}
\caption{Shapes observed in experiments and computed from simulations for $3.5\leq W/H \leq 10$ and for $L/W=1$ (first column), $L/W=3$ (second column). Numbering of the samples conforms to the measurements of the curvatures reported in Table~\ref{tab:curvatures}. The scale bar corresponds to 10\,mm.}
\label{fig:shapes2}
\end{figure}

In the computational and experimental investigation of how the aspect ratios $W/H$ and $L/W$ affect the equilibrium configuration of the bilayer, we have observed a variety of shapes that we classify as follows. \textit{Spherical caps} are samples that have almost identical, non-zero principal curvatures. \textit{Tubes}/\textit{cylindrical caps} and \textit{rings} are shapes characterized by having only one  principal curvature of appreciable magnitude along the width and the length, respectively. \textit{Barrels} are analogues of rings: they mainly bend along the length, but have a lower curvature along the width, as well. Finally, we call \textit{bananas} the samples with a predominant curvature along the width and a smaller but appreciable curvature along the length. A gallery of these shapes is illustrated in Figures~\ref{fig:shapes1}-\ref{fig:shapes2}, which demonstrate the qualitative agreement between experiments and simulations. The observed and computed shapes show a prominent feature: below a certain threshold of $W/H$ located between $10$ and $14$, samples exhibit only one type of configuration, \textit{i.e.}~spherical caps turning into barrels as the bilayer becomes wider (or equivalently thinner), while above such a threshold they can deform either into bananas or rings, with bananas evolving into cylindrical caps or tubes as $W/H$ increases. Beyond the threshold, square samples mainly bend along one of their two principal axes producing two set of shapes that may be arbitrarily distinguished into bananas/tubes or barrels/rings.
{\color{black}We remark that, as in \cite{pezzulla2016}, we also observed square samples bending along their diagonals. These shapes are not reported in the figures, as we decided to focus the present study on shapes that are symmetric with respect to the $x$ and $y$ axes.}

A quantitative characterization of the observed shapes and of the transition behavior came from the measurement of the average curvatures $\kappa_{L}$ and $\kappa_{W}$ along the length and the width, respectively, of the samples depicted in Figures~\ref{fig:shapes1}-\ref{fig:shapes2}. As regards the samples from the experiments, their average curvatures were measured from calibrated pictures taken with a digital camera Canon EOS~6D equipped with an objective EF~24-105\,mm f/4L IS USM. The experimental and numerical results are summarized in Table~\ref{tab:curvatures} and show good agreement over a wide range of aspect ratios, thus validating our theoretical model.  
Approaches based on plate theory have rationalized the transition as the result of a trade-off between bending and stretching energies: as the bilayer becomes thinner (wider in the present study), stretching energy dominates and is thus minimized almost to zero while paying for increasing but less expensive bending energy \citep{sharon2010}. %However, a quantitative comparison between theory and experiments over a wide range of aspect ratios and a classification of the shapes, especially near the transition, have not been reported previously.

\begin{figure}[!t]
\centering
\includegraphics[scale=0.29]{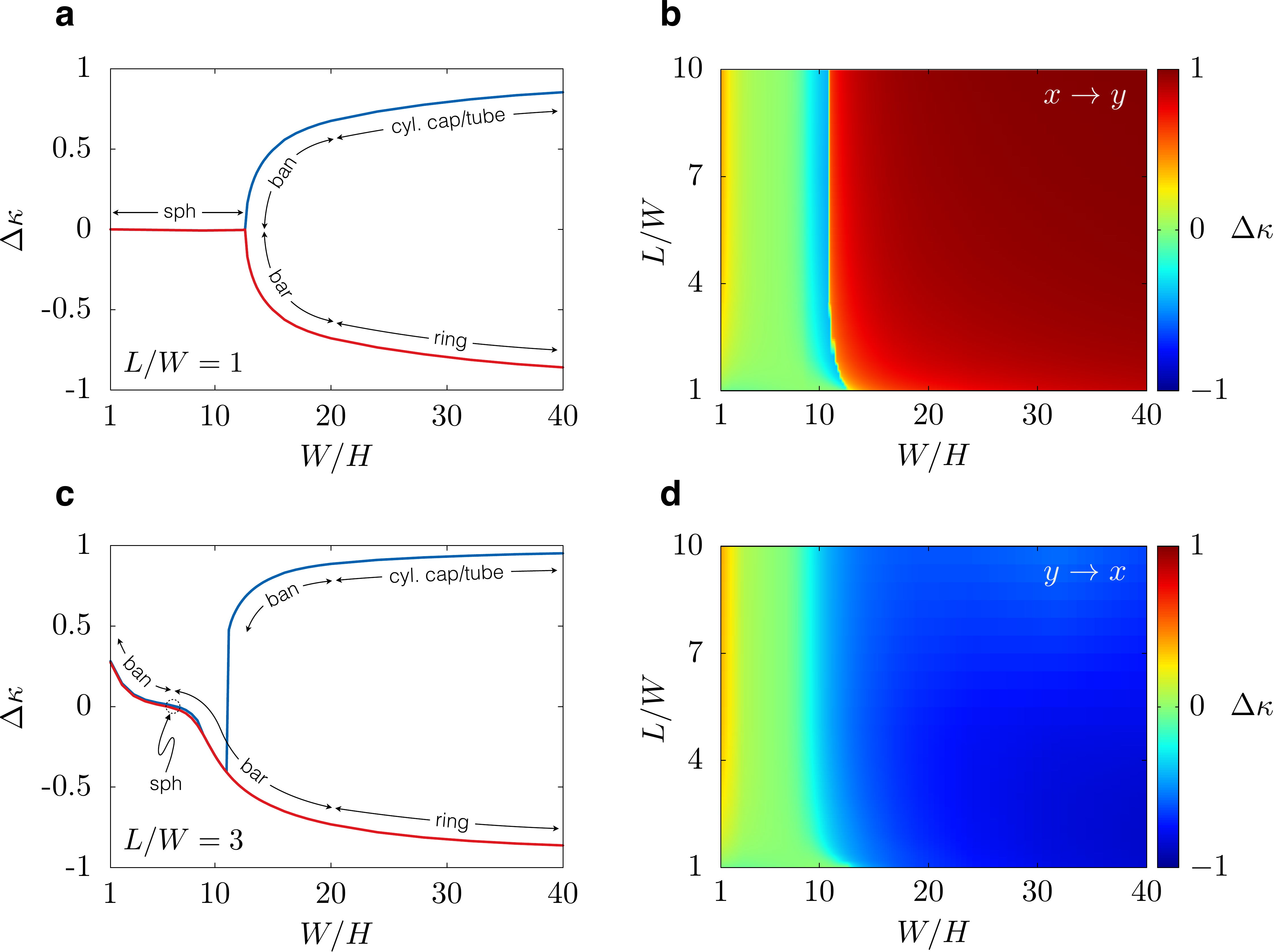}
\caption{Normalized curvature difference $\Delta \kappa$ as a function of $W/H$ for square (a, $L/W=1$) and rectangular (c, $L/W=3$) samples. The various shapes are indicated as: sph = spherical cap, ban = banana, bar = barrel, cyl.cap = cylindrical cap. Normalized curvature difference $\Delta \kappa$ as a function of $W/H$ and $L/W$, as obtained from numerical simulations following two continuation paths for the pre-stretch: increase of pre-stretch along $x$ and then along $y$ (b) and vice-versa (d). The two paths $x\rightarrow y$ and $y\rightarrow x$ correspond, respectively, to the blue and red post-bifurcation branches in (a) and (c).}
\label{fig:simulations}
\end{figure}

To examine the transition more closely, we have represented the normalized curvature difference $\Delta \kappa = (\kappa_{W} - \kappa_{L})/(\kappa_{W} + \kappa_{L})$ computed from simulations as a function of $W/H$, for typical square ($L/W=1$) and rectangular ($L/W=3$) samples (Figure~\ref{fig:simulations}a,c). In terms of the normalized curvature difference, \(\Delta \kappa = 0\) indicates a sphere, whereas the upper and lower bounds $\Delta \kappa = \pm 1$ indicate a `perfect' tube ($\kappa_L = 0$) or a `perfect' ring ($\kappa_W = 0$), respectively.  Further, $\Delta \kappa > 0$ corresponds to bananas/tubes, while $\Delta \kappa < 0$ corresponds to barrels/rings. 
The transition occurs at $W/H \simeq 11.2$ for $L/W=3$ and at $W/H \simeq 12.8$ for $L/W=1$, and is marked by a bifurcation of the principal curvatures. While such a bifurcation is continuous for squares, it shows peculiar features for rectangular samples. First, the equilibrium path barrel$\rightarrow$ring is smooth across the bifurcation, while the path barrel$\rightarrow$banana$\rightarrow$cylindrical cap$\rightarrow$tube is discontinuous. 
We believe this difference may correlate with the fact that rings are energetically preferred configurations as opposed to bananas or tubes. Indeed, as discussed in \cite{alben2011}, while both the ring and the banana/tube are stable equilibria, the former has a lower energy than the latter due to boundary layers, where the energy density is reduced. More precisely, the ring, being bent along the longer direction, will release more of the  energy stored along the edges than the tube, thus leading to a more favorable energy state. Furthermore, we notice that pre-bifurcation square samples are spherical independently of $W/H$, while rectangular samples continuously decrease the normalized curvature difference in progressing from bananas to barrels through spherical configurations. This behavior may derive from the different structure of the boundary layers of squares and rectangles, which may have a crucial impact on the stiffness of the system \citep{levin2016}.

\begin{table}[!t]
{\footnotesize
%\centering
\begin{center}
\begin{tabular}{cccccccccc}
\toprule
 \# & $W/H$ & $L/W$ & shape & $\kappa_W $ (num.) & $\kappa_W$ (exp.)  & $\delta_W$  & $\kappa_L$ (num.)  & $\kappa_L$ (exp.) & $\delta_L$ \\ \midrule
 1 & 43 & 1 & tube & 219.3 & 228.3 & 4.1\% & 12.1 & n/a & n/a \\ 
 2 & 43 & 3 & tube & 216.3 & 214.7 & 0.8\% & 3.1 & n/a & n/a \\ 
 3 & 43 & 3 & ring & 12.6 & n/a & n/a & 198.6 & 209.0 & 3.7\% \\ 
 \hline
 4 & 21.5 & 1 & cyl.cap & 206.6 & 215.2 & 4.2\% & 19.0 & n/a & n/a \\ 
 5 & 21.5 & 3 & cyl.cap & 223.7 & 201.7 & 9.8\% & 7.2 & n/a & n/a \\ 
 6 & 21.5 & 3 & ring & 24.9 & n/a & n/a & 206.9 & 182.3 & 11.9\% \\ 
 \hline
 7 & 14 & 1 & banana & 174.9 & 177.7 & 1.6\% & 61.6 & 64.2 & 4.2\%  \\ 
 8 & 14 & 3 & banana  & 217.5  & 199.4 & 8.3\% & 23.4 & 20.3 & 13.1\%  \\ 
 9 & 14 & 3 & barrel & 50.5 & 46.6 & 7.7\% & 192.6 & 203.4 & 5.6\% \\
 \hline
 10 & 10 & 1 & sph.cap & 134.9 & 119.3 & 11.6\% & 134.9 & 128.0 & 5.1\% \\ 
 11 & 10 & 3 & barrel & 70.0 & 72.5 & 3.6\% & 162.2 & 179.1 & 10.4\% \\ 
 \hline 
 12 & 7 & 1 & sph.cap & 155.4 & 123.7 & 20.4\% & 155.4 & 152.0 & 2.2\% \\ 
 13 & 7 & 3 & sph.cap & 130.6 & 124.2 & 4.9\% & 138.0 & 129.7 & 6.0\% \\ 
 \hline
 14 & 3.5 & 1 & sph.cap & 198.0 & 188.0 & 5.1\% & 198.0 & 190.0 & 4.0\% \\
 15 & 3.5 & 3 & banana & 170.0 & 162.8 & 4.2\% & 152.1 & 162.8 & 7.1\% \\ 
 \bottomrule 
\end{tabular}
\end{center} 
}
\caption{Comparison between experimental and numerical values of the average curvatures along the width, $\kappa_W$, and along the length, $\kappa_L$, of the samples, for different values of the aspect ratios $L/W$ and $W/H$ (for $L/W=1$ the identification of length and width is, of course, arbitrary). All curvatures are in m$^{-1}$. Relative differences $\delta_W$ and $\delta_L$ are computed with respect to the numerical values. Experimental values of the smallest average principal curvatures for almost developable samples showing concentrated curvature at the edges (\textit{i.e.}~cyl.~caps/tubes and rings) are omitted (n/a), due to limited reproducibility of the measurements as given by our method for curvature reconstruction.}
\label{tab:curvatures}
\end{table}

Another interesting feature that emerges from the analysis of the numerical results in Figure~\ref{fig:simulations}b,d is a shift in the location of the critical value of $W/H$ as $L/W$ varies. For $L/W<3$ the critical value of $W/H$ increases as $L/W$ decreases, while for $L/W>3$ the bifurcation threshold is independent of $L/W$. As a result, for squares the bifurcation occurs for a wider object as compared to rectangles. Indeed, square samples will less likely bifurcate into bananas/tubes since, differently from rectangular samples, they lack a preferential direction for energy reduction through boundary layers. This delay of the onset of the transition also manifests itself when a pre-stretch mismatch triggers the bifurcation, as reported in \cite{alben2011}. 

%Finally, as the width and thickness become comparable ($W/H \rightarrow 1$), the normalized curvature difference increases for rectangular samples. For square samples, we expect the edge effects to dominate much of the solid. Furthermore, the solid is no longer a thin structure similar to a plate. 

%% file: bilayers_conclusions.tex
\section{Conclusions and outlook}
\label{sec:conclusions}

In this work, we have explored the equilibrium shapes of an elastic bilayer with one of the layers pre-stretched equibiaxially, by combining experiments and numerical simulations. In particular, we have analyzed the effect of sample geometry on the principal curvatures and classified such shapes according to their average curvatures. In particular, for a fixed value of pre-stretch, making the bilayer wider (or equivalently thinner) determines a bifurcation of the principal curvatures, as the system transitions between bending-dominated and stretching-dominated regimes. We believe that our study can shed some additional light on the shape changes that occur in bilayers made of active materials, a phenomenon of interest in the framework of shape programming applications.
Future work includes a detailed analysis of the structure and energetics of boundary layers, which could be carried out using plate or beam models. Moreover, a comprehensive exploration of the shapes exhibited by square samples other than the $x$-$y$ symmetric ones considered here could be object of future studies.

%% file: bilayers_main.bbl
\begin{thebibliography}{34}
	\expandafter\ifx\csname natexlab\endcsname\relax\def\natexlab#1{#1}\fi
	\providecommand{\url}[1]{\texttt{#1}}
	\providecommand{\href}[2]{#2}
	\providecommand{\path}[1]{#1}
	\providecommand{\DOIprefix}{doi:}
	\providecommand{\ArXivprefix}{arXiv:}
	\providecommand{\URLprefix}{URL: }
	\providecommand{\Pubmedprefix}{pmid:}
	\providecommand{\doi}[1]{\href{http://dx.doi.org/#1}{\path{#1}}}
	\providecommand{\Pubmed}[1]{\href{pmid:#1}{\path{#1}}}
	\providecommand{\bibinfo}[2]{#2}
	\ifx\xfnm\relax \def\xfnm[#1]{\unskip,\space#1}\fi
	%Type = Article
	\bibitem[{Agostiniani and DeSimone(2017a)}]{agostiniani2017}
	\bibinfo{author}{Agostiniani, V.}, \bibinfo{author}{DeSimone, A.},
	\bibinfo{year}{2017}a.
	\newblock \bibinfo{title}{Dimension reduction via gamma-convergence for soft
		active materials}.
	\newblock \bibinfo{journal}{Meccanica.} \URLprefix
	\url{http://dx.doi.org/10.1007/s11012-017-0630-4},
	\DOIprefix\doi{10.1007/s11012-017-0630-4}.
	%Type = Article
	\bibitem[{Agostiniani and DeSimone(2017b)}]{agostiniani}
	\bibinfo{author}{Agostiniani, V.}, \bibinfo{author}{DeSimone, A.},
	\bibinfo{year}{2017}b.
	\newblock \bibinfo{title}{Rigorous derivation of active plate models for thin
		sheets of nematic elastomers}.
	\newblock \bibinfo{journal}{Mathematics and Mechanics of Solids\!\!} ,
	\bibinfo{pages}{1081286517699991.\,\,}\URLprefix
	\url{http://dx.doi.org/10.1177/1081286517699991},
	\DOIprefix\doi{10.1177/1081286517699991}.
	%Type = Article
	\bibitem[{Agostiniani et~al.(2016)Agostiniani, DeSimone and
		Koumatos}]{Ago_DeS_Kou}
	\bibinfo{author}{Agostiniani, V.}, \bibinfo{author}{DeSimone, A.},
	\bibinfo{author}{Koumatos, K.}, \bibinfo{year}{2016}.
	\newblock \bibinfo{title}{Shape programming for narrow ribbons of nematic
		elastomers}.
	\newblock \bibinfo{journal}{Journal of Elasticity} \bibinfo{volume}{127},
	\bibinfo{pages}{1–24}.
	\newblock \URLprefix \url{http://dx.doi.org/10.1007/s10659-016-9594-1},
	\DOIprefix\doi{10.1007/s10659-016-9594-1}.
	%Type = Article
	\bibitem[{Aharoni et~al.(2014)Aharoni, Sharon and Kupferman}]{Aha_Sha_Kup}
	\bibinfo{author}{Aharoni, H.}, \bibinfo{author}{Sharon, E.},
	\bibinfo{author}{Kupferman, R.}, \bibinfo{year}{2014}.
	\newblock \bibinfo{title}{Geometry of thin nematic elastomer sheets}.
	\newblock \bibinfo{journal}{Phys. Rev. Lett.} \bibinfo{volume}{113},
	\bibinfo{pages}{257801}.
	\newblock \URLprefix
	\url{http://link.aps.org/doi/10.1103/PhysRevLett.113.257801},
	\DOIprefix\doi{10.1103/PhysRevLett.113.257801}.
	%Type = Article
	\bibitem[{Alben et~al.(2011)Alben, Balakrisnan and Smela}]{alben2011}
	\bibinfo{author}{Alben, S.}, \bibinfo{author}{Balakrisnan, B.},
	\bibinfo{author}{Smela, E.}, \bibinfo{year}{2011}.
	\newblock \bibinfo{title}{Edge effects determine the direction of bilayer
		bending}.
	\newblock \bibinfo{journal}{Nano Letters} \bibinfo{volume}{11},
	\bibinfo{pages}{2280--2285}.
	\newblock \URLprefix \url{http://dx.doi.org/10.1021/nl200473p},
	\DOIprefix\doi{10.1021/nl200473p}.
	%Type = Article
	\bibitem[{Arroyo and DeSimone(2014)}]{arroyo2014}
	\bibinfo{author}{Arroyo, M.}, \bibinfo{author}{DeSimone, A.},
	\bibinfo{year}{2014}.
	\newblock \bibinfo{title}{Shape control of active surfaces inspired by the
		movement of euglenids}.
	\newblock \bibinfo{journal}{Journal of the Mechanics and Physics of Solids}
	\bibinfo{volume}{62}, \bibinfo{pages}{99--112}.
	\newblock \URLprefix \url{http://dx.doi.org/10.1016/j.jmps.2013.09.017},
	\DOIprefix\doi{10.1016/j.jmps.2013.09.017}.
	%Type = Article
	\bibitem[{Bartels et~al.(2017)Bartels, Bonito and Nochetto}]{Bartels}
	\bibinfo{author}{Bartels, S.}, \bibinfo{author}{Bonito, A.},
	\bibinfo{author}{Nochetto, R.H.}, \bibinfo{year}{2017}.
	\newblock \bibinfo{title}{Bilayer plates: model reduction,
		{$\Gamma$}-convergent finite element approximation, and discrete gradient
		flow}.
	\newblock \bibinfo{journal}{Communications on Pure and Applied Mathematics}
	\bibinfo{volume}{70}, \bibinfo{pages}{547–589}.
	\newblock \URLprefix \url{http://dx.doi.org/10.1002/cpa.21626},
	\DOIprefix\doi{10.1002/cpa.21626}.
	%Type = Article
	\bibitem[{Bhattacharya et~al.(1999)Bhattacharya, DeSimone, Hane, James and
		Palmstr{\o}m}]{Bhat_1999}
	\bibinfo{author}{Bhattacharya, K.}, \bibinfo{author}{DeSimone, A.},
	\bibinfo{author}{Hane, K.F.}, \bibinfo{author}{James, R.D.},
	\bibinfo{author}{Palmstr{\o}m, C.J.}, \bibinfo{year}{1999}.
	\newblock \bibinfo{title}{Tents and tunnels on martensitic films}.
	\newblock \bibinfo{journal}{Materials Science and Engineering A}
	\bibinfo{volume}{273--275}, \bibinfo{pages}{685--689}.
	\newblock \URLprefix \url{http://dx.doi.org/10.1016/S0921-5093(99)00397-4},
	\DOIprefix\doi{10.1016/S0921-5093(99)00397-4}.
	%Type = Book
	\bibitem[{Bonet and Wood(1997)}]{bonetwood}
	\bibinfo{author}{Bonet, J.}, \bibinfo{author}{Wood, R.D.},
	\bibinfo{year}{1997}.
	\newblock \bibinfo{title}{Nonlinear Continuum Mechanics for Finite Element
		Analysis}.
	\newblock \bibinfo{publisher}{Cambridge University Press}.
	%Type = Article
	\bibitem[{Chen et~al.(2011)Chen, Majidi, Sroloviz and Haataja}]{chen2011}
	\bibinfo{author}{Chen, Z.}, \bibinfo{author}{Majidi, C.},
	\bibinfo{author}{Sroloviz, D.}, \bibinfo{author}{Haataja, M.},
	\bibinfo{year}{2011}.
	\newblock \bibinfo{title}{Tunable helical ribbons}.
	\newblock \bibinfo{journal}{Applied Physics Letters,} \bibinfo{volume}{98},
	\bibinfo{pages}{011906--1:4}.
	\newblock \URLprefix \url{http://dx.doi.org/10.1063/1.3530441},
	\DOIprefix\doi{10.1063/1.3530441}.
	%Type = Article
	\bibitem[{Danescu et~al.(2013)Danescu, Chevalier, Grenet, Regreny, Letarte and
		Leclercq}]{Danescu_2013}
	\bibinfo{author}{Danescu, A.}, \bibinfo{author}{Chevalier, C.},
	\bibinfo{author}{Grenet, G.}, \bibinfo{author}{Regreny, P.},
	\bibinfo{author}{Letarte, X.}, \bibinfo{author}{Leclercq, J.},
	\bibinfo{year}{2013}.
	\newblock \bibinfo{title}{Spherical curves design for micro-origami using
		intrinsic stress relaxation}.
	\newblock \bibinfo{journal}{Applied Physics Letters} ,
	\bibinfo{pages}{123111–1:4}\URLprefix
	\url{http://dx.doi.org/10.1063/1.4798835}, \DOIprefix\doi{10.1063/1.4798835}.
	%Type = Article
	\bibitem[{DeSimone(2017)}]{desimone2017}
	\bibinfo{author}{DeSimone, A.}, \bibinfo{year}{2017}.
	\newblock \bibinfo{title}{Spontaneous bending of pre-stretched bilayers}.
	\newblock \bibinfo{journal}{Preprint\!\!} .
	%Type = Article
	\bibitem[{DeSimone et~al.(2015)DeSimone, Gidoni and Noselli}]{gidoni15}
	\bibinfo{author}{DeSimone, A.}, \bibinfo{author}{Gidoni, P.},
	\bibinfo{author}{Noselli, G.}, \bibinfo{year}{2015}.
	\newblock \bibinfo{title}{Liquid crystal elastomer strips as soft crawlers}.
	\newblock \bibinfo{journal}{Journal of the Mechanics and Physics of Solids}
	\bibinfo{volume}{84}, \bibinfo{pages}{254--272}.
	\newblock \URLprefix \url{https://doi.org/10.1016/j.jmps.2015.07.017},
	\DOIprefix\doi{10.1364/JOSA.11.000233}.
	%Type = Article
	\bibitem[{DeSimone et~al.(2013)DeSimone, Guarnieri, Noselli and
		Tatone}]{tatone13}
	\bibinfo{author}{DeSimone, A.}, \bibinfo{author}{Guarnieri, F.},
	\bibinfo{author}{Noselli, G.}, \bibinfo{author}{Tatone, A.},
	\bibinfo{year}{2013}.
	\newblock \bibinfo{title}{Crawlers in viscous environments: linear vs
		non-linear rheology}.
	\newblock \bibinfo{journal}{International Journal of Non-Linear Mechanics}
	\bibinfo{volume}{56}, \bibinfo{pages}{142–147}.
	\newblock \URLprefix \url{http://doi.org/10.1016/j.ijnonlinmec.2013.02.007},
	\DOIprefix\doi{10.1016/j.ijnonlinmec.2013.02.007}.
	%Type = Article
	\bibitem[{Freund(2000)}]{freund2000}
	\bibinfo{author}{Freund, L.}, \bibinfo{year}{2000}.
	\newblock \bibinfo{title}{Substrate curvature due to thin film mismatch strain
		in the nonlinear deformation range}.
	\newblock \bibinfo{journal}{Journal of the Mechanics and Physics of Solids}
	\bibinfo{volume}{48}, \bibinfo{pages}{1159--1174}.
	\newblock \URLprefix \url{https://doi.org/10.1016/S0022-5096(99)00070-8},
	\DOIprefix\doi{10.1016/S0022-5096(99)00070-8}.
	%Type = Article
	\bibitem[{Gioia et~al.(2002)Gioia, DeSimone, Ortiz and Cuitino}]{gioia2002}
	\bibinfo{author}{Gioia, G.}, \bibinfo{author}{DeSimone, A.},
	\bibinfo{author}{Ortiz, M.}, \bibinfo{author}{Cuitino, A.},
	\bibinfo{year}{2002}.
	\newblock \bibinfo{title}{Folding energetics in thin-film diaphragms}.
	\newblock \bibinfo{journal}{Proceedings of the Royal Society A}
	\bibinfo{volume}{458}, \bibinfo{pages}{119--192}.
	\newblock \URLprefix \url{https://doi.org/10.1098/rspa.2001.0921},
	\DOIprefix\doi{10.1098/rspa.2001.0921}.
	%Type = Article
	\bibitem[{Gioia and Ortiz(1997)}]{gioia97}
	\bibinfo{author}{Gioia, G.}, \bibinfo{author}{Ortiz, M.}, \bibinfo{year}{1997}.
	\newblock \bibinfo{title}{Delamination of compressed thin films}.
	\newblock \bibinfo{journal}{Advances in Applied Mechanics}
	\bibinfo{volume}{33}, \bibinfo{pages}{1223--1229}.
	\newblock \URLprefix \url{http://doi.org/10.1016/S0065-2156(08)70386-7},
	\DOIprefix\doi{S0065-2156(08)70386-7}.
	%Type = Article
	\bibitem[{Guest et~al.(2011)Guest, Kebadze and Pellegrino}]{guest2011}
	\bibinfo{author}{Guest, S.}, \bibinfo{author}{Kebadze, E.},
	\bibinfo{author}{Pellegrino, S.}, \bibinfo{year}{2011}.
	\newblock \bibinfo{title}{A zero-stiffness elastic shell structure}.
	\newblock \bibinfo{journal}{Journal of Mechanics of Materials and Structures}
	\bibinfo{volume}{6}, \bibinfo{pages}{203--212}.
	\newblock \URLprefix \url{http://dx.doi.org/10.2140/jomms.2011.6.203},
	\DOIprefix\doi{jomms.2011.6.203}.
	%Type = Article
	\bibitem[{Guest and Pellegrino(2006)}]{guest2006}
	\bibinfo{author}{Guest, S.}, \bibinfo{author}{Pellegrino, S.},
	\bibinfo{year}{2006}.
	\newblock \bibinfo{title}{Analytical models for bistable cylindrical shells}.
	\newblock \bibinfo{journal}{Proceedings of the Royal Society A}
	\bibinfo{volume}{462}, \bibinfo{pages}{839--854}.
	\newblock \URLprefix \url{http://dx.doi.org/10.2140/10.1098/rspa.2005.1598},
	\DOIprefix\doi{10.1098/rspa.2005.1598}.
	%Type = Article
	\bibitem[{Huang et~al.(2012)Huang, Liu, Kroll, Bertoldi and Clarke}]{huang2012}
	\bibinfo{author}{Huang, J.}, \bibinfo{author}{Liu, J.}, \bibinfo{author}{Kroll,
		B.}, \bibinfo{author}{Bertoldi, K.}, \bibinfo{author}{Clarke, D.},
	\bibinfo{year}{2012}.
	\newblock \bibinfo{title}{Spontaneous and deterministic three-dimensional
		curling of pre-strained elastomeric bi-strips}.
	\newblock \bibinfo{journal}{Soft Matter} \bibinfo{volume}{8},
	\bibinfo{pages}{6291--6300}.
	\newblock \URLprefix \url{http://dx.doi.org/10.1039/C2SM25278C},
	\DOIprefix\doi{10.1039/C2SM25278C}.
	%Type = Article
	\bibitem[{Kim et~al.(2012)Kim, Hanna, Byun, Santangelo and Hayward}]{kim2012}
	\bibinfo{author}{Kim, J.}, \bibinfo{author}{Hanna, J.A.},
	\bibinfo{author}{Byun, M.}, \bibinfo{author}{Santangelo, C.D.},
	\bibinfo{author}{Hayward, R.C.}, \bibinfo{year}{2012}.
	\newblock \bibinfo{title}{Designing responsive buckled surfaces by halftone gel
		lithography}.
	\newblock \bibinfo{journal}{Science} \bibinfo{volume}{335},
	\bibinfo{pages}{1201--1205}.
	\newblock \URLprefix \url{http://science.sciencemag.org/content/335/6073/1201},
	\DOIprefix\doi{10.1126/science.1215309}.
	%Type = Article
	\bibitem[{Klein et~al.(2007)Klein, Efrati and Sharon}]{klein2007}
	\bibinfo{author}{Klein, Y.}, \bibinfo{author}{Efrati, E.},
	\bibinfo{author}{Sharon, E.}, \bibinfo{year}{2007}.
	\newblock \bibinfo{title}{Shaping of elastic sheets by prescription of
		non-euclidean metrics}.
	\newblock \bibinfo{journal}{Science,} \bibinfo{volume}{315},
	\bibinfo{pages}{1116--1120}.
	\newblock \URLprefix \url{http://dx.doi.org/10.1126/science.1135994},
	\DOIprefix\doi{10.1126/science.1135994}.
	%Type = Article
	\bibitem[{Levin and Sharon(2016)}]{levin2016}
	\bibinfo{author}{Levin, I.}, \bibinfo{author}{Sharon, E.},
	\bibinfo{year}{2016}.
	\newblock \bibinfo{title}{Anomalously {Soft} {Non}-{Euclidean} {Springs}}.
	\newblock \bibinfo{journal}{Physical Review Letters} \bibinfo{volume}{116},
	\bibinfo{pages}{035502}.
	\newblock \URLprefix \url{http://dx.doi.org/10.1103/PhysRevLett.116.035502},
	\DOIprefix\doi{10.1103/PhysRevLett.116.035502}.
	%Type = Article
	\bibitem[{Lucantonio et~al.(2014)Lucantonio, Nardinocchi and
		Pezzulla}]{lucantonio2014}
	\bibinfo{author}{Lucantonio, A.}, \bibinfo{author}{Nardinocchi, P.},
	\bibinfo{author}{Pezzulla, M.}, \bibinfo{year}{2014}.
	\newblock \bibinfo{title}{Swelling-induced and controlled curving in layered
		gel beams}.
	\newblock \bibinfo{journal}{Proceedings of the Royal Society A}
	\bibinfo{volume}{470}, \bibinfo{pages}{20140467}.
	\newblock \URLprefix
	\url{http://rspa.royalsocietypublishing.org/cgi/doi/10.1098/rspa.2014.0467},
	\DOIprefix\doi{10.1098/rspa.2014.0467}.
	%Type = Article
	\bibitem[{Lucantonio et~al.(2017)Lucantonio, Tomassetti and
		DeSimone}]{lucantonio2016}
	\bibinfo{author}{Lucantonio, A.}, \bibinfo{author}{Tomassetti, G.},
	\bibinfo{author}{DeSimone, A.}, \bibinfo{year}{2017}.
	\newblock \bibinfo{title}{Large-strain poroelastic plate theory for polymer
		gels with applications to swelling-induced morphing of composite plates}.
	\newblock \bibinfo{journal}{Composites Part B: Engineering}
	\bibinfo{volume}{115}, \bibinfo{pages}{330--340}.
	\newblock \URLprefix \url{https://doi.org/10.1016/j.compositesb.2016.09.063},
	\DOIprefix\doi{10.1016/j.compositesb.2016.09.063}.
	%Type = Article
	\bibitem[{Mansfield(1962)}]{mansfield1962}
	\bibinfo{author}{Mansfield, E.H.}, \bibinfo{year}{1962}.
	\newblock \bibinfo{title}{Bending, {Buckling} and {Curling} of a {Heated}
		{Thin} {Plate}}.
	\newblock \bibinfo{journal}{Proceedings of the Royal Society A}
	\bibinfo{volume}{268}, \bibinfo{pages}{316--327}.
	\newblock \URLprefix
	\url{http://rspa.royalsocietypublishing.org/content/268/1334/316},
	\DOIprefix\doi{10.1098/rspa.1962.0143}.
	%Type = Article
	\bibitem[{Modes et~al.(2011)Modes, Bhattacharya and Warner}]{Bhat_Mod_War}
	\bibinfo{author}{Modes, C.D.}, \bibinfo{author}{Bhattacharya, K.},
	\bibinfo{author}{Warner, M.}, \bibinfo{year}{2011}.
	\newblock \bibinfo{title}{Gaussian curvature from flat elastica sheets}.
	\newblock \bibinfo{journal}{Proceedings of the Royal Society A}
	\bibinfo{volume}{467}, \bibinfo{pages}{1121--1140}.
	\newblock \URLprefix \url{http://dx.doi.org/10.1098/rspa.2010.0352},
	\DOIprefix\doi{10.1098/rspa.2010.0352}.
	%Type = Article
	\bibitem[{Mostajeran(2015)}]{Mostajeran}
	\bibinfo{author}{Mostajeran, C.}, \bibinfo{year}{2015}.
	\newblock \bibinfo{title}{Curvature generation in nematic surfaces}.
	\newblock \bibinfo{journal}{Phys. Rev. E} \bibinfo{volume}{91},
	\bibinfo{pages}{062405}.
	\newblock \URLprefix \url{http://link.aps.org/doi/10.1103/PhysRevE.91.062405},
	\DOIprefix\doi{10.1103/PhysRevE.91.062405}.
	%Type = Article
	\bibitem[{Na et~al.(2015)Na, Evans, Bae, Chiappelli, Santangelo, Lang, Hull and
		Hayward}]{na2015}
	\bibinfo{author}{Na, J.H.}, \bibinfo{author}{Evans, A.A.},
	\bibinfo{author}{Bae, J.}, \bibinfo{author}{Chiappelli, M.C.},
	\bibinfo{author}{Santangelo, C.D.}, \bibinfo{author}{Lang, R.J.},
	\bibinfo{author}{Hull, T.C.}, \bibinfo{author}{Hayward, R.C.},
	\bibinfo{year}{2015}.
	\newblock \bibinfo{title}{Programming reversibly self-folding origami with
		micropatterned photo-crosslinkable polymer trilayers}.
	\newblock \bibinfo{journal}{Advanced Materials} \bibinfo{volume}{27},
	\bibinfo{pages}{79--85}.
	\newblock \URLprefix
	\url{http://onlinelibrary.wiley.com/doi/10.1002/adma.201403510/abstract},
	\DOIprefix\doi{10.1002/adma.201403510}.
	%Type = Article
	\bibitem[{Noselli and DeSimone(2014)}]{noselli14}
	\bibinfo{author}{Noselli, G.}, \bibinfo{author}{DeSimone, A.},
	\bibinfo{year}{2014}.
	\newblock \bibinfo{title}{A robotic crawler exploiting directional frictional
		interactions: experiments, numerics, and derivation of a reduced model}.
	\newblock \bibinfo{journal}{Proceedings of the Royal Society A}
	\bibinfo{volume}{470}, \bibinfo{pages}{20140333}.
	\newblock \URLprefix \url{http://doi.org/10.1098/rspa.2014.0333},
	\DOIprefix\doi{10.1098/rspa.2014.0333}.
	%Type = Article
	\bibitem[{Pezzulla et~al.(2015)Pezzulla, Shillig, Nardinocchi and
		Holmes}]{pezzulla2015}
	\bibinfo{author}{Pezzulla, M.}, \bibinfo{author}{Shillig, S.A.},
	\bibinfo{author}{Nardinocchi, P.}, \bibinfo{author}{Holmes, D.P.},
	\bibinfo{year}{2015}.
	\newblock \bibinfo{title}{Morphing of geometric composites via residual
		swelling}.
	\newblock \bibinfo{journal}{Soft Matter} \bibinfo{volume}{11},
	\bibinfo{pages}{5812--5820}.
	\newblock \URLprefix \url{http://dx.doi.org/10.1039/C5SM00863H},
	\DOIprefix\doi{10.1039/C5SM00863H}.
	%Type = Article
	\bibitem[{Pezzulla et~al.(2016)Pezzulla, Smith, Nardinocchi and
		Holmes}]{pezzulla2016}
	\bibinfo{author}{Pezzulla, M.}, \bibinfo{author}{Smith, G.P.},
	\bibinfo{author}{Nardinocchi, P.}, \bibinfo{author}{Holmes, D.P.},
	\bibinfo{year}{2016}.
	\newblock \bibinfo{title}{Geometry and mechanics of thin growing bilayers}.
	\newblock \bibinfo{journal}{Soft Matter} \bibinfo{volume}{12},
	\bibinfo{pages}{4435--4442}.
	\newblock \URLprefix \url{http://dx.doi.org/10.1039/C6SM00246C},
	\DOIprefix\doi{10.1039/C6SM00246C}.
	%Type = Article
	\bibitem[{Sharon and Efrati(2010)}]{sharon2010}
	\bibinfo{author}{Sharon, E.}, \bibinfo{author}{Efrati, E.},
	\bibinfo{year}{2010}.
	\newblock \bibinfo{title}{The mechanics of non-{Euclidean} plates}.
	\newblock \bibinfo{journal}{Soft Matter} \bibinfo{volume}{6},
	\bibinfo{pages}{5693}.
	\newblock \URLprefix \url{http://dx.doi.org/10.1039/c0sm00479k},
	\DOIprefix\doi{10.1063/1.4798835}.
	%Type = Article
	\bibitem[{Timoshenko(1925)}]{timoshenko25}
	\bibinfo{author}{Timoshenko, S.}, \bibinfo{year}{1925}.
	\newblock \bibinfo{title}{Analysis of bi-metal thermostats}.
	\newblock \bibinfo{journal}{J. Optical Soc. Am.,} \bibinfo{volume}{11},
	\bibinfo{pages}{233--255}.
	\newblock \URLprefix \url{http://doi.org/10.1364/JOSA.11.000233},
	\DOIprefix\doi{10.1364/JOSA.11.000233}.
	
\end{thebibliography}
